\documentstyle[prd,aps,twocolumn]{revtex}
\newcommand{\barxi}{\overline\xi}
\begin{document}
\title{Gravitational clustering in a $D$-dimensional Universe.}
\author{T. Padmanabhan$^1$\footnote{paddy@iucaa.ernet.in} \& Nissim Kanekar$^2$\footnote{nissim@ncra.tifr.res.in}}
\address{$1$ Inter-University Centre for Astronomy and Astrophysics,
 Post Bag 4, Ganeshkhind, Pune 411 007, India}
\address{$2$ National Centre for Radio Astrophysics,
 Post Bag 3, Ganeshkhind, Pune 411 007, India}
\maketitle
\begin{abstract}
\noindent We consider the problem of gravitational clustering in a $D$-dimensional 
expanding Universe and derive scaling relations connecting the exact mean 
two-point correlation function with the linear mean correlation function, 
in the quasi-linear and non-linear regimes, using the standard paradigms of 
scale-invariant radial collapse and stable clustering.
We show that the existence of scaling laws is a generic feature of 
gravitational clustering in an expanding background, in all dimensions 
except $D=2$ and comment on the special nature of the 2-dimensional 
case. The $D$-dimensional scaling laws derived here reduce, in the 3-dimensional 
case, to scaling relations obtained earlier from $N$-body simulations. 
Finally, we consider the case of clustering of 2-dimensional particles 
in a 2-$D$ expanding background, governed by a force $-GM/R$, and show that 
the correlation function does not grow (to first order) until much after 
the recollapse of any shell.\\
\end{abstract} 
%\vskip -0.7 in
\begin{center}
\vskip -0.12 in
{\large {\bf I. Introduction}}
\end{center}
%\section{Introduction }
\vskip -0.12 in
\noindent The temporal evolution of a system consisting of a large number of 
particles interacting with each other via Newtonian gravity is a formidable 
problem to tackle. Such a system has, in fact, no ``final'' state 
of thermodynamic equilibrium as there is no limit to its phase space volume. 
The phase volume available can be continuously increased by the 
separation of particles into a collapsed core and a dispersed halo, with 
the core becoming more and more tightly bound and the halo particles moving 
to larger and larger distances. In fact, the only stable configuration of 
such a system consists of a tightly coupled binary, with all the other 
particles at infinite distance.\\
\noindent The above situation changes drastically if the background of the 
system ({\it i.e.} the space in which the particles are moving) is itself 
expanding. In this case, the expansion provides a ``stabilising'' influence 
on the evolution and can result in the formation of stable structures in 
which the effects of gravitational collapse are balanced by the expansion. 
Further, there also exists the possibility that the outmoving halos may 
be captured by other compact cores, leading to the build up of larger 
structures. Thus, the issue of gravitational clustering of a large number 
of particles appears more tractable in the case of a background expanding 
(in general) with a time-dependant scale factor. This problem is, of course, 
of  immense physical relevance as there currently exists strong evidence 
that the matter density in the Universe 
is dominated by collisionless dark matter particles, which interact 
solely by gravity. Thus, if the length scales of interest are much smaller 
than the Hubble scale (and particle velocities are non-relativistic),
the formation of large-scale structure in the Universe is well described by 
the above picture, making it worthy of investigation.\\
\noindent In the linear regime, where deviations from uniformity are 
small, perturbative techniques are used to obtain the time evolution of the 
system parameters, for example, the correlation functions. These methods, 
of course, fail in the quasi-linear and non-linear regimes and there is, as 
yet, no clear analytical picture of the behaviour of the system in 
these stages. However, in the case of 3+1 dimensions, there exists a set of
scaling relations relating the linear and non-linear correlation functions 
(\cite{ham}, \cite{paddy}). These scaling laws are {\it not} 
particularly well understood but appear to be validated by numerical simulations 
(\cite{ham}, \cite{pad95}). \\
\noindent A better understanding 
of a physical problem is sometimes attained by treating it in a more general manner 
in an arbitrary number of dimensions as one may then be able to separate 
the generic features of the problem (as in, issues arising from the nature 
of the interaction) from results which stem from its dimensionality. This 
has been seen earlier, for example, in the case of the Ising model. In the 
current letter, we address the issue of gravitational clustering 
in a $D$-dimensional, expanding Universe and attempt to derive 
the scaling relations, using the well-known paradigms of scale-invariant 
radial infall in the quasi-linear phase and stable clustering in the non-linear 
regime.\\
%\vskip -0.4 in
\begin{center}
\vskip -0.12 in
{\large {\bf II. Clustering in $D$-dimensions}}
\end{center}
%\section{Clustering in $D$-dimensions}
\vskip -0.12 in
\noindent We consider the case of a $D$-dimensional
Universe, expanding with a time-dependant scale factor $a(t)$. Further, we 
use the maximally symmetric Robertson-Walker metric (in $D+1$ dimensions) 
and specialise to flat space ($k = 0$). The equations governing the evolution 
of $a(t)$ are then (see \cite{esp}) \\
\begin{equation}
\label{frw1}
\frac{\dot{a}^2}{a^2} = \frac{2\kappa(D)}{D(D-1)} \rho_b 
\end{equation}
\begin{equation}
\label{frw2}
\frac{\ddot{a}}{a} + \frac{D-2}{2} \left(\frac{\dot{a}}{a}\right)^2 = 
-\frac{\kappa(D)}{(D-1)}p
\end{equation}
\noindent where $\rho_b$ and $p$ are, respectively, the density and 
pressure of the background Universe and $\kappa$ is the constant in 
the Einstein equations, which, in general, can depend on the dimension $D$. 
We note that equations (\ref{frw1}) and (\ref{frw2}) are obtained by imposing 
the constraints of homogeneity and isotropy on the Einstein equations 
in $(D+1)$ dimensions; the situation is thus manifestly isotropic. 
Further, it can be seen that the behaviour in the $D = 2$ case is likely 
to be special, due to the presence of the $(D-2)$ coefficient in the second term 
of equation (\ref{frw2}). We emphasise that the present work considers 
the clustering of $D$-dimensional particles in a $D$-dimensional expanding 
Universe. Earlier studies of 2-$D$ clustering in the literature (see, for 
example, \cite{bep}) treated the clustering of infinite needles in a 3-dimensional 
background; this will be commented upon later. \\ 
\noindent Given an equation of state, $p(\rho_b)$, one can 
solve equations (\ref{frw1}) and (\ref{frw2}) for the evolution of $a(t)$. 
For pressureless dust, with $p = 0$, this implies that $\rho_b$ is given by \\
\begin{equation}
\rho_b \propto a^{-D}
\end{equation}
\noindent Next, we incorporate the effects of perturbations on the smooth 
background by considering the evolution of the above system starting from 
Gaussian initial conditions with an initial power spectrum $P_{in}(k)$. 
The two-point correlation function, $\xi(a,x)$, is defined as the Fourier 
transform of the power spectrum. We will, for convenience, work 
with the {\it mean} two-point correlation function, $\barxi(a,x)$, defined by \\
\begin{equation}
\barxi(a,x) = {D \over {x^D}} \int_0^x dy y^{D-1} \xi(a,y)
\end{equation}
\noindent and attempt to relate the exact $\barxi(a,x)$ in the quasi-linear and 
non-linear regimes to the mean correlation function calculated from linear theory, 
at the same epoch. \\
\noindent The equation for the conservation of pairs can be written in 
$D$-dimensions as \\
\begin{equation}
{{\partial \xi} \over {\partial t}} + {1\over{ax^{D-1}}}{{\partial\over{\partial{x}}}}
\left[  x^{D-1}v(1+\xi)\right] = 0
\end{equation} 
\noindent where $v(a,x)$ is the mean relative pair velocity at scale $x$ and time $t$. 
In terms of $\barxi(a,x)$, this gives \\
\begin{equation}
\left[ {\partial \over {\partial A}} - h(a,x) {\partial \over {\partial X}} \right]
{\rm ln } (1 + \barxi) = D h(a,x)
\end{equation}
\noindent Here, $h(a,x) = - v/{\dot a}x$, $X = \mbox {ln } x $ and $A = 
\mbox{ln } a$. The above equation can be further simplified by defining 
$F = {\rm ln}\left[ x^D (1 + \barxi ) \right]$, yielding \\
\begin{equation}
\label{char}
\left[ {\partial \over {\partial A}} - h(a,x){\partial\over{\partial X}} \right]F= 0
\end{equation}
\noindent The characteristic curves of this equation, on which $F$ is a 
constant, satisfy the condition ${\rm ln}\left[ x^D (1 + \barxi ) \right]  = 
{\rm constant} $, {\it i.e. } 
\begin{equation}
\label{lxeqn}
x^D (1 + \barxi ) = l^D
\end{equation}
\noindent where $l$ is some other length scale. (Note that $x^D (1 + \barxi )$ is 
proportional to the number of neighbours of a given particle, within a 
$D$-dimensional sphere of radius $ax$; the above equation expresses 
the conservation of pairs in this sphere \cite{lss}.) When the evolution is linear at
all relevant scales, $\barxi(a,x) \ll 1$ and $x \approx l$. However, in the 
non-linear regime, $\barxi(a,x) \gg 1$ at some scale $x$; clearly $x \ll l$. Thus, 
the behaviour of clustering at a scale $x$ is determined by the transfer of 
power from a larger scale $l$ along the characteristics defined by equation 
(\ref{char}). This suggests that one should try to relate the true $\barxi(a,x)$ 
to the linear correlation function $\barxi_L(a,l)$, evaluated at a {\it different} 
scale.\\
\noindent The paradigm of scale-invariant radial collapse (see \cite{fg84}, \cite{bert}) 
will be used 
to carry out the above procedure, in the quasi-linear regime. Consider the evolution 
of a $D$-dimensional, spherically symmetric, overdense region, containing 
a mass $M$.  In general, such a region will initially expand with the 
background Universe until the excess gravitational force due 
to its enclosed mass causes it to collapse back upon itself. Thus, the radius 
of the region will initially rise, reach a maximum and then decrease.
The equation of motion for the radius, $R(t)$, of such a $D$-dimensional, 
spherical region is (\cite{esp})\\
\begin{equation}
\label{reqn}
{{d^2 R}\over {dt^2}} = - {{2(D-2)}\over {D^2}} (1 + \delta_i) {{l^2} \over
{t_i^2}} {{l^{D-2}} \over {R^{D-1}}}
\end{equation}
\noindent where we have replaced for the mass, $M$, in terms of $\delta_i$, 
$l$ and $t_i$, {\it i.e.} the initial density contrast, shell radius and time, 
respectively. We note that the acceleration, $(d^2R/dt^2)$, is proportional 
to $1/R^{D-1}$, in $D$ dimensions. The energy integral for equation (\ref{reqn}) is \\
\begin{equation}
E = {1 \over 2} \left( {{dR} \over {dt}}\right)^2 - {{2(1+\delta_i)}\over {D^2}}
{{l^2} \over {t_i^2}} \left( {{l} \over R} \right)^{D-2}
\end{equation}
\noindent The above expression is, of course, not valid for $D = 2$; this 
case will be treated separately later. We evaluate the energy constant, $E$, by 
requiring that the velocity satisfies the unperturbed expansion at $t = t_i$, 
{\it i.e.} $dR/dT = l H_i = 2 l/ Dt_i$, where $H_i = 2/D t_i$ is the Hubble parameter. 
This gives \\
\begin{equation}
\left( {{dR} \over {dt}}\right)^2 = {{4(1+\delta_i)}\over {D^2}}
{{l^2} \over {t_i^2}} \left( {{l} \over R} \right)^{D-2} - {{4\delta_i}
\over {D^2}} {{l^2} \over {t_i^2}}
\end{equation}
\noindent At turnaround, $R = R_M$ and $dR/dt = 0$. Thus \\
\begin{equation}
\left( {{\delta_i} \over {1 + \delta_i}} \right) = \left( {{l}\over 
{R_M}}\right)^{D-2}
\end{equation}
\noindent or, since $\delta_i \ll 1$, \\
\begin{equation}
R_M =  l \delta_i^{1/(2-D)}
\end{equation}
\noindent In the quasi-linear regime, we expect clustering to take place 
in a region surrounding density peaks of the linear regime, {\it i.e.} around 
regions such as the spherical region considered above. Making the usual assumption 
that the typical density profile around such a peak is equal to the average profile 
around a mass point, we can write this density profile as \\
\begin{equation}
\rho(x) \approx \rho_{\rm b}(1 + \barxi)
\end{equation}
\noindent Thus, the initial density contrast, $\delta_i(l) \propto \barxi_L(l)$, 
in the initial epoch, when linear theory is valid. Since $R_M = 
l \delta_i^{1/(2-D)}$, clearly $R_M \propto \barxi_L^{1/(2-D)}$. 
In the scale-invariant, 
radial collapse picture, each shell can be approximated as contributing an 
effective radius proportional to $R_M$. Taking the final effective radius, 
$x$, as proportional to $R_M$, the final mean correlation function is given by \\
\begin{equation}
\barxi(x) \propto \rho \propto {M \over {x^D}} \propto {{l^D} \over {\left[ l 
\barxi_L^{1/(2-D)}\right]^D}} \propto 
\left[ \barxi_L(l)\right]^{D/(D-2)}
\end{equation}
\noindent Thus, the final correlation function $\barxi_{\rm QL}$ at $x$ is the 
$D/(D-2)$ power of the initial correlation function at $l$, where $l^D \propto x^D
\left[ \barxi_L(l)\right]^{{D/(D-2)}} \propto x^D \barxi_{\rm QL}(x)$, which 
is the form required by equation (\ref{lxeqn}), if $\barxi_{\rm QL} \gg 1$. \\
\noindent Let us turn our attention, next, to the non-linear regime; here, we 
use the ansatz of stable clustering 
($h \rightarrow 1$ for $\barxi \gg 1$), which is physically well-motivated as it 
seems reasonable to expect stable, bound systems to form under the joint influence 
of gravity and the expansion (\cite{lss}). Such systems would neither expand nor 
contract and would hence have peculiar velocities equal and opposite to the Hubble 
expansion, {\it i.e. } $v^i = -{\dot a}x^i$. (We emphasise that the stable clustering 
hypothesis is an {\it ansatz} and might well fail if mergers of structures are 
important.) Stable clustering requires that virialized systems 
should maintain their densities and sizes in {\it proper} co-ordinates, $r = ax$.
 This would require the correlation function, in $D$ dimensions, to have the form 
$\barxi_{\rm NL}(a,x) = a^D F(ax)$, where $F$ is some function and the $a^D$ 
factor arises from the decrease in the background density. We assume that 
$\barxi_{\rm NL}(a,x)$ is a function of $\barxi_L(a,l)$, where $l^D \approx 
x^D \barxi_{\rm NL}(a,x)$, from equation (\ref{lxeqn}). This relation can be 
written as \\
\begin{equation}
\barxi_{\rm NL}(a,x) = a^D F(ax) = U\left[ \barxi_L(a,l) \right] 
\end{equation}
\noindent where $U(p)$ is an unknown function of its argument, which is to be 
determined. The density contrast and mean correlation function have the 
relation $\barxi \propto \delta^2$, in the linear regime; this 
arises from the definition of $\xi$ as the Fourier transform of $\delta^2$. 
Now, in a flat $D$-dimensional matter-dominated Universe, $\delta \propto 
a^{D-2}$ (\cite{esp}), in the linear regime. Thus, $\barxi_L \propto 
\delta_l^2 \propto a^{2(D-2)}$. We can hence write $\barxi_L(a,l) = 
a^{2(D-2)} Q\left[ l^D \right]$, where $Q$ is a function of $l$ alone. But, 
$l^D \approx  x^D \barxi_{\rm NL}(a,x) = x^D a^D F(ax) = r^D F(r)$. This implies \\
\begin{eqnarray}
\barxi_{\rm NL}(a,x)&=&a^D F(r) = U\left[ \barxi_L(a,l) \right]\\
&=& U\left[ a^{2(D-2)} Q \left[ r^D F(r) \right]   \right] 
\end{eqnarray}
\noindent Considering the above relation as a function of $a$ at constant $r$, we 
obtain  \\
\begin{equation}
Ba^D = F\left[ C a^{2(D-2)} \right]
\end{equation}
\noindent where $B$ and $C$ are constants. One must hence have \\
\begin{equation}
F(p) \propto p^{D/2(D-2)}
\end{equation}
\noindent Thus, in the extreme non-linear regime, we obtain \\
\begin{equation}
\barxi_{\rm NL}(a,x) \propto \left[ \barxi_L(a,l) 
\right]^{D/2(D-2)}
\end{equation}
\noindent The above analysis shows that the exact mean correlation function 
can be expressed in terms of the linear mean correlation function by the relation \\
\begin{equation}
\barxi(a,x) = \barxi_L(a,l)       \hskip 1.1 in (\mbox {linear})
\end{equation}
\begin{equation}
\barxi(a,x)\propto \left[ \barxi_L(a,l) \right]^{D/(D-2)} 
\hskip 0.1 in (\mbox {quasi-linear})
\end{equation}
\begin{equation}
\barxi(a,x)    \propto \left[ \barxi_L(a,l) \right]^{D/2(D-2)} 
\hskip 0.1 in (\mbox {non-linear})
\end{equation}
\noindent In the case of 3 dimensions, the above equations reduce to $\barxi(a,x) \propto 
\left[ \barxi_L(a,l) \right]^3$ and $\left[ \barxi_L(a,l) \right]^{3/2}$ in the 
quasi-linear and non-linear regimes, respectively; these are in reasonable agreement 
with simulations (\cite{ham}, \cite{pad95}; see, however, \cite{peacock}). 
Thus, scaling laws appear to be a generic feature of gravitational clustering in an 
expanding background, in all dimensions (except $D=2$). \\
\noindent The 2-dimensional case is special and different from all other dimensions, 
as, in this case, equation (\ref{reqn}) gives \\
\begin{equation}
{{d^2R} \over{dt^2}} = 0
\end{equation}
\noindent {\it i.e.} the correlation function does not evolve at all in 2-$D$ and no 
structures are formed. The special nature of the $D=2$ case appears to arise from 
the structure of the Poisson equation, due to the fact that it contains derivatives 
of the second order. It has been noted earlier that scaling relations 
{\it can} arise in 2-$D$ numerical simulations of gravitational clustering 
(\cite{bep}). These simulations, however, do not consider the 
case of 2-$D$ particles evolving in a 2-$D$ background but instead treat the system 
as consisting of a set of infinitely long, parallel needles, in a 3-$D$ expanding 
background, with particles considered as arising at the intersections of the 
needles with any plane orthogonal to them. The mass elements in the needles interact 
via the usual $1/{r^2}$ force; however, the interaction between the needles themselves 
is governed by a $1/r$ force. The simulations consider the clustering of the needles 
by taking a slice through a plane orthogonal to them. This situation is, of course, 
clearly {\it anisotropic} as clustering is considered as taking place in 2 dimensions 
while the background Universe expands in 3 dimensions. In this case, one can show 
numerically (\cite{fg84}) that $R_M = l/\delta_i$ in the quasi-linear 
regime and arrive at the relevant scaling relations by an analysis similar to the above. \\
\noindent We finally consider the case of 2-$D$ particles in a 2-$D$ 
expanding background and {\it assume} a force law \\
\begin{equation}
\label{2eqn}
{{d^2R} \over{dt^2}} = -{{GM} \over R}
\end{equation} 
\noindent The above form, of course, {\it cannot} be obtained in a 
self-consistent manner, by taking limits of the Einstein equations 
for an expanding Universe; however, it has the form $d^2R/dt^2 \propto 
R^{-1}$, the correct dependence for $D=2$ in equation (\ref{reqn}).
\noindent The energy integral of equation (\ref{2eqn}) is \\
\begin{equation}
E = \frac{1}{2} \left( \frac {dR}{dt}\right)^2 - GM \;\mbox{ln}R
\end{equation}
\noindent The mass $M$, enclosed within the shell, is given by \\
\begin{equation}
M = (1 + \delta_i) \rho_b \pi l^2
\end{equation}
\noindent where $l$ is the initial shell radius. Here, we will assume 
$\rho_b \propto t^{-2}$, the usual result for $D$ dimensions. Setting 
$dR/dt = H_i l$ initially, and using $H_i \propto 1/t_i$, we obtain \\
\begin{equation}
R_M = l\; \mbox{exp}\left[ -A(1 + \delta_i)\right]
\end{equation}
\noindent where $A$ is a constant; {\it i.e.} $R_M/l = const.$, to leading 
order in $\delta_i$. Again taking the 
final effective radius, of the 2-dimensional shell, as proportional to $R_M$, 
the final mean correlation function is given by \\
\begin{equation}
\overline\xi(x) \propto \rho \propto \frac{M}{x^2} \propto \frac{l^2}{l^2} 
\propto \mbox{const.}
\end{equation}
\noindent Thus, the correlation function does not grow (to first order) 
until much after turnaround. This is an interesting result which needs to 
be verified by simulations.\\
\noindent The above situation is similar to the case of a 3-$D$ network of parallel 
cosmic strings (space around them is flat and $\rho_b \propto a^{-2}$). 
Structure will clearly {\it not} grow if the network is expanding uniformly;
however, if the strings had some initial peculiar velocities, 
then two strings which pass on either side of a third would move towards 
each other because the angle around the third string is less than $2\pi$.
\begin{center}
\vskip -0.12 in
{\large {\bf III. Conclusions}}
\end{center}
%\section{Conclusions}
\vskip -0.12 in
\noindent In the present paper, the standard paradigms of scale-invariant radial 
collapse and stable clustering have been used to derive scaling relations 
connecting the exact mean two-point correlation function and the linear 
mean correlation function, for a $D$-dimensional expanding Universe. 
The existence of scaling laws is found to be a generic feature of gravitational 
clustering in an expanding background, in all dimensions, except $D=2$. Further, 
the scaling laws derived here are in agreement, in the 3-dimensional 
case, with scaling relations obtained earlier from $N$-body simulations 
($\bar\xi \propto \bar\xi_{L}^3$ and $\bar\xi \propto \bar\xi_{L}^{3/2}$, 
in the quasi-linear and non-linear 
regimes, respectively.). Finally, we have considered the case of clustering of 
2-$D$ particles in a 2-dimensional expanding background, governed by a force -GM/R, 
and show, by a similar analysis, that the correlation function does not grow, to 
first order, until much after turnaround of any shell. \\

\end{document}